\documentclass[aps,floatfix,twocolumn,showpacs,preprintnumbers]{revtex4}

\usepackage{mathrsfs}
\usepackage{amsmath, amssymb, amsfonts, bbm}
\usepackage{graphics,epsfig,color,subfig, graphicx}

\usepackage{natbib}

\newcommand{\be}{\begin{equation}}
\newcommand{\ee}{\end{equation}}
\def\bea{\begin{align}}
\def\ena{\end{align}}

\def\beqa{\begin{eqnarray}}
\def\enqa{\end{eqnarray}}

\newcommand{\N}{N_c}

\begin{document}

\title{Alternate 1/$N_c$ Expansions and SU(3) Breaking from Baryon Lattice
Results}

\author{Aleksey Cherman}
\email{a.cherman@damtp.cam.ac.uk}
\affiliation{Department of Applied Mathematics and Theoretical Physics,
Cambridge University, Cambridge CB3 0WB, UK}

\author{Thomas D. Cohen}
\email{cohen@physics.umd.edu}
\affiliation{Maryland Center for Fundamental Physics, Department of
Physics, University of Maryland, College Park, Maryland 20742-4111, USA}

\author{Richard F. Lebed}
\email{Richard.Lebed@asu.edu}
\affiliation{Department of Physics, Arizona State University, Tempe,
Arizona 85287-1504, USA}

\date{May 2012}

\begin{abstract}
A combined expansion in the number of QCD colors $1/N_c$ and SU(3)
flavor-breaking parameter $\epsilon$ has long been known to provide an
excellent accounting for the mass spectrum of the lightest spin-$\frac
{1}{2} , \frac{3}{2}$ baryons when the quarks are taken to transform
under the fundamental SU($N_c$) representation, and in the final step
$N_c \to 3$ and $\epsilon$ is set to its physical value $\sim 0.3$.
Subsequent work shows that placing quarks in the two-index
antisymmetric SU($N_c$) representation leads to quantitatively equally
successful mass relations.  Recent lattice simulations allow for
varying the value of $\epsilon$ and confirm the robustness of the
original $1/N_c$ relations.  In this paper we show that the same
conclusion holds for the antisymmetric quarks, and demonstrate that
the mass relations also hold under alternate prescriptions for
identifying physical baryons with particular members of the large
$N_c$ multiplets.
\end{abstract}


%

\maketitle

\section{Introduction}
QCD with three colors and realistic quark masses has no expansion
parameters to allow perturbative calculations of low-energy
observables from first principles.  Soon after the discovery of QCD,
however, 't~Hooft observed~\cite{tHooft:1973jz} that non-Abelian gauge
theories such as QCD simplify in the limit that the number of colors
$N_{c}$ tends to infinity, suggesting that one could try to compute
observables in an expansion in the small parameter $1/N_{c}$.  The
notion of large $N$ limits has been extraordinarily fruitful for
studies of the formal aspects of gauge theories, and many of the
qualitative predictions of the $1/N_{c}$ expansion are reasonable when
compared to experimental results.

Unfortunately, because of the lack of a solution of large $N_{c}$ QCD
in four dimensions, only a limited number of quantitative predictions
has been obtained from the $1/N_{c}$ expansion.  Despite some initial
skepticism~\cite{Coleman:1980nk} about the phenomenological relevance
of the properties of baryons at large $N_{c}$~\cite{Witten:1979kh},
these successes have turned out to involve primarily large $N_{c}$
baryons rather than mesons.  The most powerful tool for extracting
such predictions has been the contracted SU$(2N_{f})$ spin-flavor
symmetry that emerges at large $N_{c}$ for
baryons~\cite{Gervais:1984rc, Dashen:1993jt,Dashen:1993as,
Dashen:1994qi,Carone:1993dz,Luty:1993fu}.  When combined with the
SU$(3)$ flavor-breaking expansion, the SU$(2N_{f})$ spin-flavor
symmetry was shown to imply relations between baryon masses, even
including isospin splittings~\cite{Jenkins:1995td,Jenkins:2000mi}, and
these relations show a good fit to the experimental baryon mass
spectra.  Indeed, these relations are among the strongest pieces of
evidence that the large $N_{c}$ expansion is necessary to understand
real-world baryon spectroscopy, since the combined $1/N_{c}$ and
flavor-breaking expansion predictions fit the data much better than
the predictions from flavor-breaking alone.  The robustness of baryon
mass relations has recently been confirmed by showing they continue to
hold to predicted uncertainties when tested on baryon mass spectra
obtained from lattice simulations~\cite{WalkerLoud:2008bp} as the size of the SU$(3)$ flavor
breaking is varied~\cite{Jenkins:2009wv}.

However, recent theoretical developments have revealed that the
predictions resulting from the SU$(2N_{f})$ symmetry are not quite
unique.  The reason is the existence of more than one
phenomenologically viable way to take a large $N_{c}$ limit starting
from $N_{c}=3$.  The limit advocated in the seminal papers of
't~Hooft~\cite{tHooft:1973jz} and Witten~\cite{Witten:1979kh} assumed
that quarks are in the fundamental (F) representation for all $N_{c}
\geq 3$, while keeping the number of flavors $N_{f}$ fixed.  We refer
to this as the {\it large} $N_{c}^{\rm F}$ {\it limit}.  However,
one can obtain a different large $N_{c}$ limit by taking the $N_{f}$
quarks to be in the two-index antisymmetric (AS) representation of
SU$(N_{c})$ for $N_{c} \geq
3$~\cite{Armoni:2003gp,Armoni:2003fb,Armoni:2004uu,Armoni:2005wt}; we
call this the {\it large\/} $N_{c}^{\rm AS}$ {\it limit}.  The reason
for two possible extrapolations to general $N_{c}$ from $N_{c}=3$ is
that the AS and F representations are isomorphic for $N_{c}=3$.

The large $N_{c}^{\rm AS}$ limit also implies the emergence of an
SU$(2N_{f})$ spin-flavor symmetry, and hence it also results in baryon
mass relation predictions, but with different $1/N_{c}$ suppression
factors than the large $N_{c}^{\rm F}$ limit.  Curiously, it turns out
that the experimental data appears to be consistent with both large
$N_{c}$ limits to within the experimental and theoretical
errors~\cite{Cherman:2009fh}.  Apparently, {\em some\/} $1/N_{c}$
expansion is necessary to fit the data, but it is not possible to
decide which large $N_{c}$ limit provides a better guide for baryon
mass spectra.  The SU$(2N_{f})$ symmetry also makes predictions for
magnetic moments of baryons, and here the large $N_{c}^{\rm F}$
expansion is a better guide to the data than the large $N_{c}^{\rm
AS}$ expansion~\cite{Lebed:2010um}.

An additional complication leads to possible ambiguities in large
$N_{c}$ predictions for baryon properties.  At large $N_{c}$, color
antisymmetrization allows baryons to have many more than 3 quarks, and
consequently many more baryon species appear than at $N_{c}=3$; one
needs a prescription to match the baryons that can exist at $N_{c}=3$
to the baryons seen at large $N_{c}$.  With more than two flavors, no
unique prescription exists, and one finds at least two natural
extrapolations of baryons with valence strange quarks at $N_{c}=3$ to
large $N_{c}$.  With one choice one obtains large $N_{c}$ baryons with
$\sim N_{c}^{1}$ strange quarks, while with the other one obtains only
$\sim N_{c}^{0}$ strange quarks.  The baryon properties are different
depending upon the prescription one adopts, and this is a serious
phenomenological challenge in comparing to data.

In this paper, we address both of these subtleties in the context of
the large $N_{c}$ baryon mass relations.  First, we use the lattice
data employed by Jenkins {\it et al.}~\cite{Jenkins:2009wv} to compare
the predictions of the large $N_{c}^{\rm F}$ and large $N_{c}^{\rm
AS}$ limits as a function of the strength of SU$(3)$ breaking.  We
find that they both remain consistent with the data.  Second, we
discuss the ambiguities in matching the $N_{c}=3$ and large
$N_{c}$ flavor representations of baryons.  We show that remarkably,
while individual baryon masses are sensitive to these ambiguities, the
baryon mass relations are not.  It is tempting to speculate that this
may in some sense be the reason for the robustness and
phenomenological success of the large $N_{c}$ baryon mass relations.

The organization of this paper is as follows.
Section~\ref{sec:Baryons} presents a brief review of the operator
methods by which large $N_c$ baryon phenomenology is performed.
Section~\ref{sec:2Expansions} presents an analysis of the baryon mass
lattice results in light of the large $N_{c}^{\rm AS}$ expansion, and
shows that this expansion remains phenomenologically just as relevant
for this observable as does the large $N_{c}^{\rm F}$ limit.  The
robustness of baryon mass relations under different prescriptions for
treating strangeness is discussed in Sec.~\ref{sec:NcandSU3}, and
Sec.~\ref{sec:Conclusions} offers concluding remarks.

\section{Baryons at Large $N_c$}
\label{sec:Baryons}

The conventional phenomenological operator analysis of large $N_c$
baryons is based upon the use of three specific properties: (i) the
large number of valence quarks in the baryon [which is $N_c$ in the
large $N_c^{\rm F}$ expansion and $N_c(N_c-1)/2$ in the large
$N_c^{\rm AS}$ expansion] and their detailed combinatorics in the
baryon wave function~\cite{Witten:1979kh,Bolognesi:2006ws}, (ii) the 't~Hooft
scaling~\cite{tHooft:1973jz} $g \! \sim \! N_c^{-1/2}$ of the QCD
coupling constant required to obtain a nontrivial large $N_c$ limit,
and (iii) a ground-state multiplet whose states are completely
symmetric under the combined spin-flavor symmetry, and whose $N_c =
3$, $N_f = 3$ case is the SU(6) {\bf
56}-plet~\cite{Dashen:1993jt,Carone:1993dz,Luty:1993fu}.  Ambiguities
related to the identification of the physical baryon states with
particular states within the large $N_c$ multiplets are addressed in
Sec.~\ref{sec:NcandSU3}.  The nonvalence (gluon and sea quark) degrees
of freedom enter into the analysis only indirectly: Since the physical
baryons fill specific spin and flavor representations based upon the
quantum numbers of valence quarks, the entire baryon wave function may
be written in terms of interpolating fields that carry the same quark
spin/flavor/color quantum numbers and that encompass the full baryon
wave function, thus providing a rigorous footing to the concept of
constituent quarks~\cite{Buchmann:2000wf}.

Operators that describe the interactions among the component quarks of
the baryon are classified in terms of their transformation properties
under the spin-flavor symmetry~\cite{Dashen:1994qi} and the number $n$
of quark fields appearing in the interaction, hence defining an {\it
$n$-body operator}.  Upon including both the appropriate quark
combinatorics and the 't~Hooft scaling, one finds that $n$-body
operators are generically suppressed by a factor $1/N_c^n$
($1/N_c^{2n}$) in the $N_c^{\rm F}$ ($N_c^{\rm AS}$) counting.
However, obtaining the full $1/N_c$ suppression factor requires one to
account for two other sources: First, one must also consider
the possibility that contributions from the quarks {\em add
coherently\/} in the baryon matrix elements, which introduces
compensatory combinatoric factors of $N_c$ that may change the power
counting.  Second, certain combinations of operators may give matrix
elements that form a linearly dependent set when evaluated on the
baryon multiplet in question; such a linear dependence may occur due to
an exact operator identity (for example, in the case of Casimir
operators) or due to the symmetry properties of a particular multiplet
on which the analysis is performed (for example, the vanishing of an
antisymmetric tensor acting upon the completely symmetric ground-state
multiplet)~\cite{Dashen:1994qi}.  In addition, a particular
combination of operators acting upon a particular multiplet might
produce a result that is subleading in the $1/N_c$ expansion compared
to each of the component operators, which is termed an {\it operator
demotion}.  Once this reorganization is complete, one is left with a
linearly independent set of operators, each one of which produces
matrix elements with a well-defined power counting in $1/N_c$ acting
upon a particular baryon multiplet, and this set carries precisely the
same dimension as the space of independent baryon observables for the
multiplet.  One sees that the operators and independent observables
form equivalent bases for the baryons, and the operators may then be
assembled into an effective Hamiltonian for any baryon observable such
that the operators form a hierarchy in powers of $1/N_c$.

To see how these properties work in detail, we begin by
defining operators with specific transformation properties under the
spin-flavor symmetry.  The operators in the adjoint representation are
labeled $J^i$, $T^a$, and $G^{ia}$, respectively, in
Refs.~\cite{Jenkins:1995td,Jenkins:2000mi,Dashen:1993jt}:
\begin{eqnarray} \label{su6}
J^i & \equiv & q^\dagger_\alpha \left( \frac{\sigma^i}{2} \otimes
\openone \right) q^\alpha , \nonumber \\
T^a & \equiv & q^\dagger_\alpha \left( \openone \otimes
\frac{\lambda^a}{2} \right) q^\alpha , \nonumber \\
G^{ia} & \equiv & q^\dagger_\alpha \left( \frac{\sigma^i}{2} \otimes
\frac{\lambda^a}{2} \right) q^\alpha ,
\end{eqnarray}
where $\openone$ is the identity matrix, $\sigma^i$ are Pauli matrices
in spin space, $\lambda^a$ is the usual Gell-Mann matrix in flavor
space, and $\alpha$ sums over all the quarks in the baryon.  Color
indices do not appear explicitly in this expression since the baryon
ground-state multiplet is completely symmetric under the combined
spin-flavor symmetry and therefore is completely antisymmetric under
color, so that the operators are well-defined for both large $N_c^{\rm
F}$ and large $N_c^{\rm AS}$.  All operators that have nonvanishing
matrix elements on baryon states are expressible as polynomials in
$J^i$, $T^a$, and $G^{ia}$ (with suitable contractions of spin-flavor
indices), and such a polynomial of $n$th degree is an $n$-body
operator.  Since the physical baryons have $N_c = 3$ quarks, any such
polynomial beyond cubic order applied to these baryons gives matrix
elements linearly dependent upon those of lower-order operators, which
means that such operators are ignorable; the $1/N_c$ series for any
finite given value of $N_c$ terminates after providing a complete set
of independent operators, which does not extend beyond the $N_c$-body
level.

An $n$-body operator requires an $n$-quark interaction, which in turn
implies $2n$ factors of the QCD coupling $g$, to give the $1/N_c^n$
suppression for the large $N_c^{\rm F}$ limit, as discussed above.
One might naively expect the same suppression factor for the large
$N_c^{\rm AS}$ case, but one finds, as argued in~\cite{Cherman:2006iy}
and systematically verified in~\cite{Cohen:2009wm}, that the necessity
of maintaining the color-singlet nature of the large $N_c^{\rm AS}$
baryon effectively makes the appropriate effective expansion parameter
$1/N_c^2$.  Therefore, in the effective large $N_c^{\rm F}$ baryon
Hamiltonian, the operator $T^a$ appears multiplied by an explicit
factor of $1/N_c$ compared to the spin-flavor symmetric operator
$\openone$ that has $O(N_c^1)$ matrix elements [while, in large
$N_c^{\rm AS}$, the corresponding factors are $1/N_c^2$ and
$O(N_c^2)$].  Consider now the mass operator $T^8$; its matrix
elements, which naively merely count strange quarks, are actually
given in large $N_c^{\rm F}$ by
\begin{equation}
\langle T^8 \rangle = \frac{1}{2\sqrt{3}}(N_c - 3N_s) \, .
\end{equation}
Here we see a coherent $O(N_c^1)$ contribution that seems to upset the
large $N_c$ counting; however, note that it is the same for all
baryons and therefore simply provides an additional contribution to
the leading-order spin-flavor symmetric mass operator $\openone$.  In
the above terminology, $T^8 - \openone /2\sqrt{3} $ is a demoted
operator.  The operator $T^8$ also breaks SU(3)$_{\rm flavor}$ and
therefore requires an explicit prefactor of $\epsilon$.  When one
repeats this analysis for a complete set of linearly independent
operators, one finds that each operator contributes to a unique baryon
mass combination.  To be specific, the mass Hamiltonian when isospin
breaking is suppressed reads~\cite{Jenkins:1995td}:
\begin{align}
M = & c_{(0)}^{1,0}\ \N \openone + c_{(2)}^{1,0}\ \frac{1}{\N}
J^2 \nonumber \\
& +  c_{(1)}^{8,0}\ \epsilon T^8 + c_{(2)}^{8,0}\ \frac{\epsilon}{\N}
\{ J^i, G^{i8} \} + c_{(3)}^{8,0}\ \frac{\epsilon}{\N^2} \{ J^2, T^8 \} 
\nonumber \\
& +   c_{(2)}^{27,0}\ \frac{\epsilon^2}{\N} \{ T^8, T^8 \}
+  c_{(3)}^{27,0}\ \frac{\epsilon^2}{\N^2} \{ T^8, \{ J^i, G^{i8} \}\} 
\nonumber \\
& +  c_{(3)}^{64,0} \frac{\epsilon^3}{\N^2} \{ T^8, \{ T^8, T^8\}\} \,
,
\label{ops} \end{align}
where the coefficients $c$ are $O(N_c^0)$, and the nontrivial matrix
elements of the baryon operators are given by
\begin{align}
& O^{8,0}_{(2)} \equiv \frac{\epsilon}{N_c} \{ J^i, G^{i8} \}
\nonumber \\
& \to \frac{1}{2\sqrt{3}} \cdot \frac{\epsilon}{N_c}
\left[ 3 I (I+1) - J(J+1)
- \frac{3N_s}{2}\left (\frac{N_s}{2} +1 \right) \right] \, ,
\nonumber \\
&  O^{27,0}_{(3)} \equiv \frac{\epsilon^2}{N_c^2}
\{ T^8, \{ J^i, G^{i8} \}\} \nonumber \\
& \to \frac{\epsilon^2}{N_c^2} \cdot \frac 1 6 (N_c \! - 3N_s)
\nonumber
\\ & \times \left[ 3 I (I + \! 1) \! - J(J \! +1)
- \! 3 \frac{N_s}{2} \! \left (\frac{N_s}{2} +1 \right) \right] ,
\nonumber \\
&  O^{8,0}_{(1)} \equiv \epsilon T^8  \to \epsilon \cdot
\frac{1}{2\sqrt{3}} \left( N_c - 3 N_s \right) \, ,
\nonumber\\
& O^{27,0}_{(2)} \equiv \frac{\epsilon^2}{N_c} \{ T^8, T^8 \} \to
\frac{1}{6} \frac{\epsilon^2}{N_c} \left( N_c - 3 N_s \right)^2 \, ,
\nonumber\\
& O^{64,0}_{(3)} \equiv \frac{\epsilon^3}{N_c^2}
\{ T^8, \{ T^8, T^8\}\} \to \frac{\epsilon^3}{N_c^2}
\frac{1}{6\sqrt{3}} \left (N_c - 3 N_s \right)^3 \; ,
\label{matrixelements}
\end{align}
where the explicit suppressions of $\epsilon$ or $1/N_c$ have here
been absorbed into the operator definitions.  The operators of
Eq.~(\ref{ops}) define the combinations $M_i$ in
Table~\ref{MassRelations}, including the $1/N_c$ and $\epsilon$
suppression factors.  For example, the combination $M_2$ is associated
with the operator $\epsilon T^8$ considered above.

\begin{center}
\begin{table*}
\caption{Baryon mass combinations and their theoretical suppression
factors.  The 0 subscript indicates an average over isospin states,
which are termed $I=0$ baryon masses.
\label{MassRelations}}
\begin{tabular}{| c | c | c | c | }
\hline
      & Mass combination & Large $N_c^{\mathrm{F}}$ suppression &
      Large $N_c^{\mathrm{AS}}$ suppression \\
\hline
$M_0$ & $25(2N_0 + 3\Sigma_0 + \Lambda +2\Xi_0) -4(4\Delta_0
+3\Sigma^*_0 +2\Xi^*_0 +\Omega)$  & $N_c$ & $N_c^2$  \\
\hline
$M_1$ &  $5(2N_0 + 3\Sigma_0 + \Lambda +2\Xi_0) -4(4\Delta_0
+3\Sigma^*_0 +2\Xi^*_0 +\Omega)$  & $1/N_c$ & $1/N_c^2$  \\
\hline
$M_2$ & $5(6N_0 -3\Sigma_0 +\Lambda -4\Xi_0) -2(2\Delta_0 -\Xi^*_0
-\Omega) $ & $\epsilon$ & $\epsilon$ \\
\hline
$M_3$ & $N_0 -3\Sigma_0 +\Lambda +\Xi_0$ & $\epsilon / N_c$ &
$\epsilon / N_c^2$ \\
\hline
$M_4$ & $ (-2N_0 -9\Sigma_0+3\Lambda + 8\Xi_0) +2(2\Delta_0 -\Xi^*_0
-\Omega)$ & $\epsilon / N_c^2$ & $\epsilon / N_c^4$ \\
\hline
$M_5$ & $35(2N_0 -\Sigma_0 -3\Lambda +2\Xi_0) -4(4\Delta_0
-5\Sigma^*_0 -2\Xi^*_0 +3\Omega)$ & $\epsilon^2 / N_c$ & $\epsilon^2 /
N_c^2$ \\
\hline
$M_6$ & $ 7 (2N_0 -\Sigma_0-3\Lambda + 2\Xi_0) -2(4\Delta_0
-5\Sigma^*_0 -2\Xi^*_0 +3\Omega) $ & $\epsilon^2 / N_c^2$ &
$\epsilon^2 / N_c^4$ \\
\hline
$M_7$ & $\Delta_0 - 3 \Sigma^*_0 + 3 \Xi^*_0 - \Omega$ & $\epsilon^3 /
N_c^2$ & $\epsilon^3 / N_c^4$ \\
\hline
\end{tabular}
\end{table*}
\end{center}

Such an analysis is not unique to the $1/N_c$ expansion; all that is
required is a finite multiplet of states under some symmetry and a
perturbative parameter that suppresses some of the independent
operators acting upon the multiplet.  For example, since an operator
with matrix elements linear in strangeness breaks SU(3)$_{\rm flavor}$
symmetry by transforming as an {\bf 8}, an operator transforming as a
{\bf 27} ($\subset {\bf 8} \otimes {\bf 8}$) does not occur until
second order in flavor breaking, and this operator must be associated
with a doubly-suppressed flavor-breaking mass combination.  Indeed,
when evaluated for the $N_c = 3$ baryon octet, this mass combination
turns out to be just the one that appears in the Gell-Mann--Okubo
relation, $2N_0 -\Sigma_0 -3\Lambda +2\Xi_0$, where $X_0$ indicates
the isospin average of the $X$ baryon isomultiplet masses.

\section{The $1/N_c^{\rm F}$ and $1/N_c^{\rm AS}$ Expansions}
\label{sec:2Expansions}

Our previous results~\cite{Cherman:2009fh} show that the predictions
of both the large $N_{c}^{\rm F}$ and large $N_{c}^{\rm AS}$ analyses
fit the experimental spectrum of $I=0$ baryon masses comparably well.
In the real world, one can take the SU(3)$_{\rm flavor}$-breaking
parameter to be, {\it e.g.}, $\epsilon = (m_K^2 -
m_\pi^2)/\Lambda_\chi^2
\approx 0.226$ ($\Lambda_\chi \approx 1$~GeV indicating the scale of
chiral symmetry breaking); however, lattice calculations of baryon
spectra allow one to move away from the physical value of $\epsilon$,
and Jenkins {\it et al.}~\cite{Jenkins:2009wv} demonstrated that the
predictions of the large $N_{c}^{\rm F}$ expansion continue to
accommodate the data well even as $\epsilon$ is varied over the range
(0,0.26).  In this section we compare the predictions of the
$1/N_{c}^{\rm AS}$ expansion to the lattice data.

First, let us briefly review how these comparisons are defined in
Refs.~\cite{Cherman:2009fh,Jenkins:2009wv}.  Each mass combination
$M_{i}$ from Table~\ref{MassRelations} (our $M_{0,\ldots,7}$
corresponding, respectively, to $M_{1,\ldots,8}$ in
Ref.~\cite{Jenkins:2009wv}) is associated with a dimensionless ratio
$R_{i} \equiv M_{i}/(M'_{i}/2)$, where $M'_{i}$ is defined to be the
same combination of masses as in $M_{i}$, but with each coefficient
replaced with its absolute value.  These ratios therefore compare the
size of $M_{i}$ as extracted from lattice simulations or experiment
relative to the appropriately weighted average $M'_{i}$ of the masses
that enter the mass combinations $M_{i}$, which makes $R_i$
scale-independent quantities (but note that our $R_i$'s differ from
those defined in~\cite{Jenkins:2009wv}).  We then compute (as
in~\cite{Cherman:2009fh}) the ratio of each $R_i$ to its corresponding
theoretical suppression $S_{i}$, including the appropriate SU(3)$_{\rm
flavor}$ and $N_{c}$ factors as listed in Table~\ref{MassRelations},
and plot finally the {\it accuracy factors\/} $A_{i} \equiv \log_{3}
(R_{i}/S_{i})$.  The logarithm appears in order to distinguish
different integer-power suppressions, while the base 3 is used to
separate each factor of $N_c = 3$ by one unit.  In particular,
deviations from $R_{i}/S_{i} \sim 1$ to either, {\it e.g.},
$R_{i}/S_{i} = 1/2$ or $R_{i}/S_{i} = 2$ are considered equally
significant.  If the suppression factors in the measured quantities
agree with those predicted theoretically, one expects that the $A_{i}$
should lie roughly in the range $-1
\lesssim A_{i} \lesssim 1$.

\begin{figure*}
  \centering
  \subfloat[$A_{0}$]{\label{fig:A1Plot}
\includegraphics[width=0.4\textwidth]{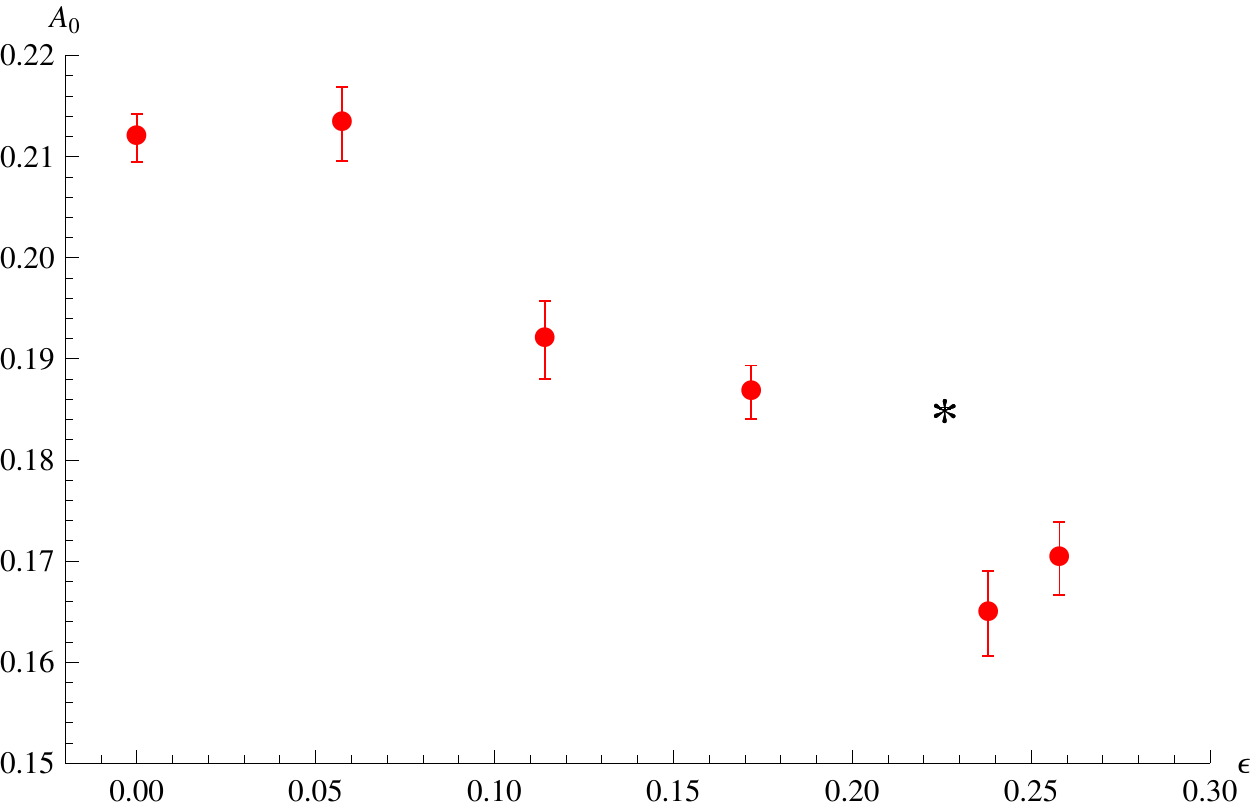}}
  \subfloat[$A_{1}$]{\label{fig:A2Plot}
\includegraphics[width=0.4\textwidth]{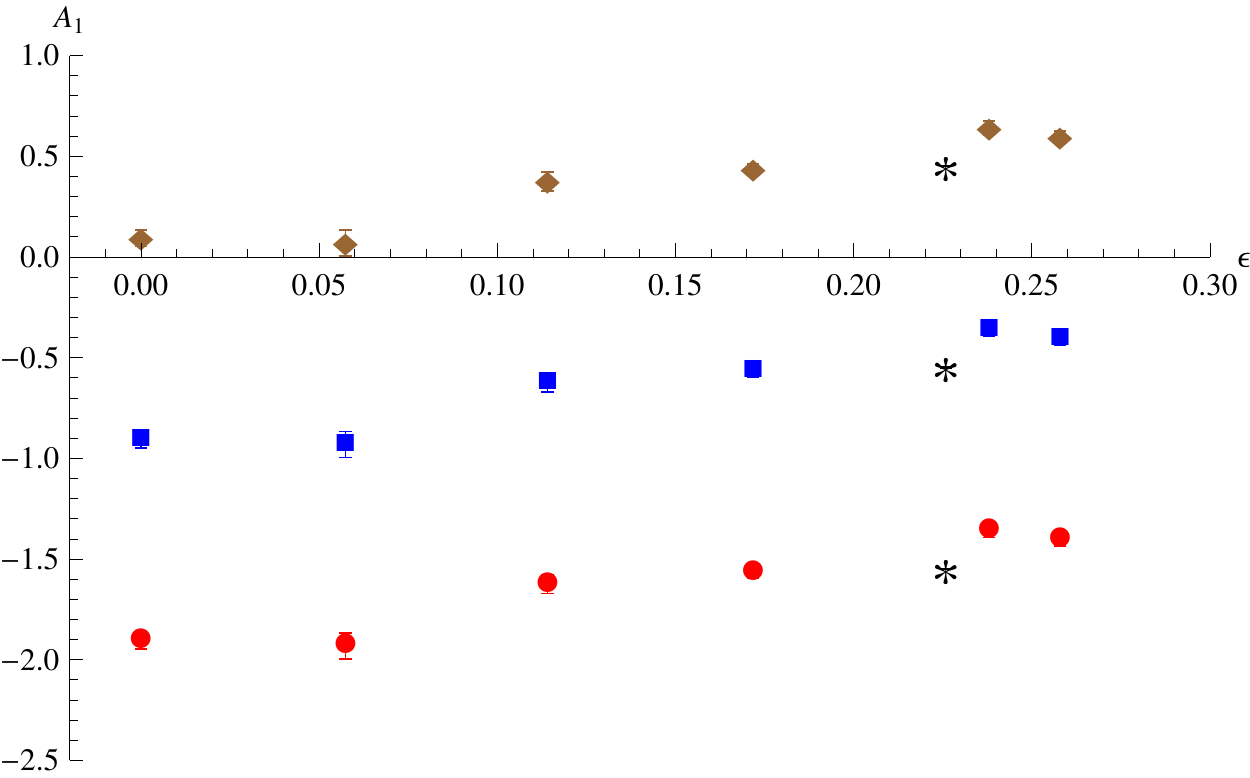}}
  \qquad
  \subfloat[$A_{2}$]{\label{fig:A3Plot}
\includegraphics[width=0.4\textwidth]{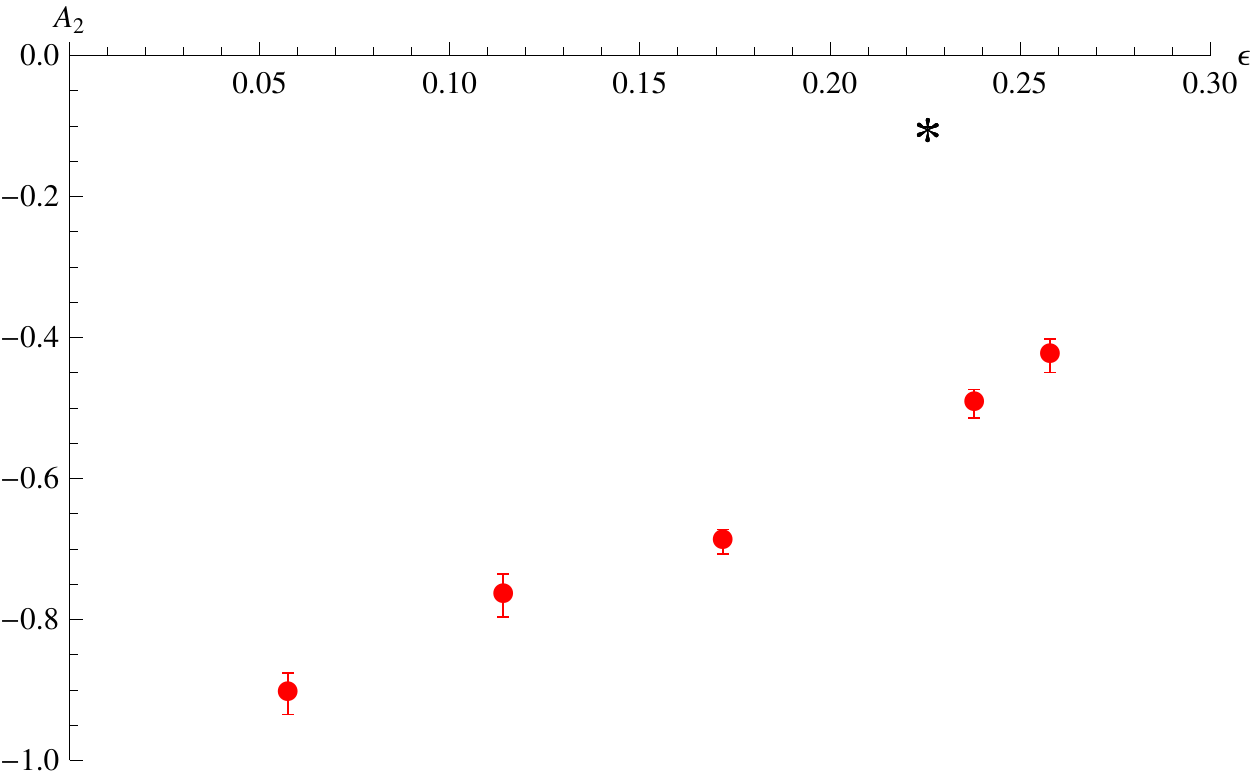}}
  \subfloat[$A_{3}$]{\label{fig:A4Plot}
\includegraphics[width=0.4\textwidth]{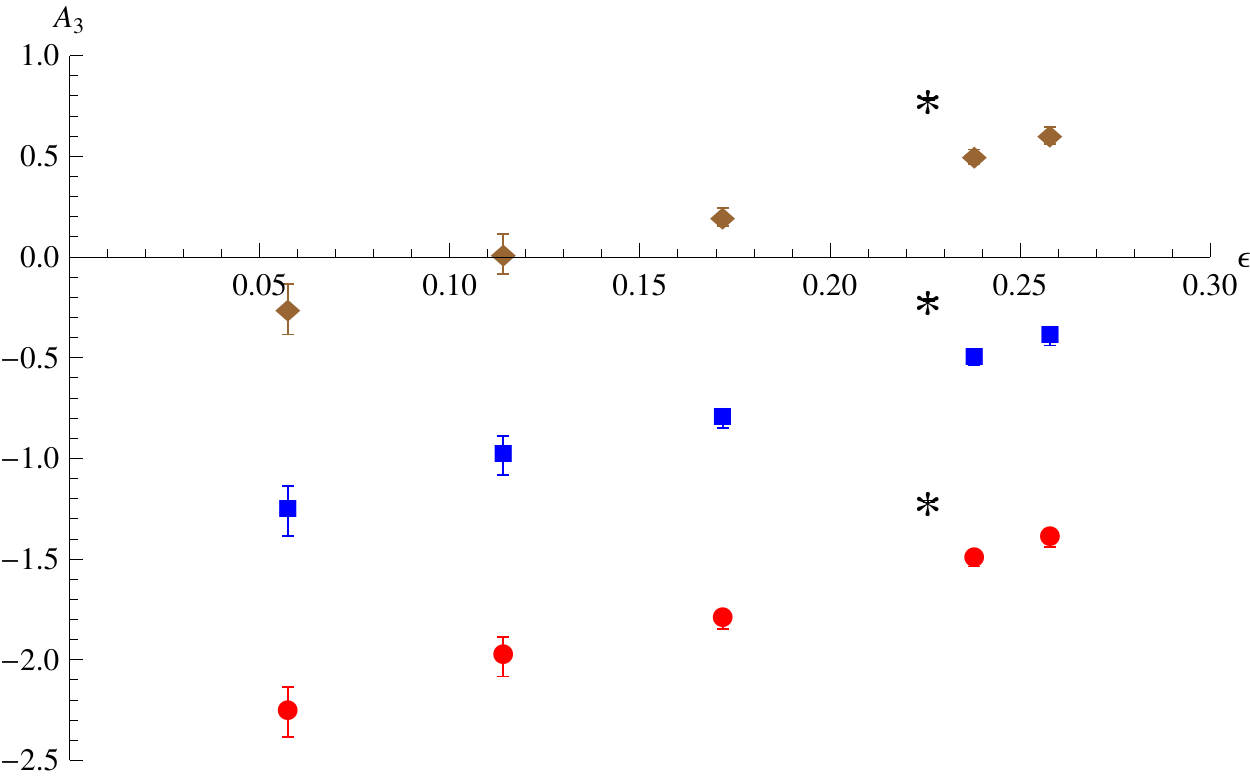}}
  \qquad
  \subfloat[$A_{4}$]{\label{fig:A5Plot}
\includegraphics[width=0.4\textwidth]{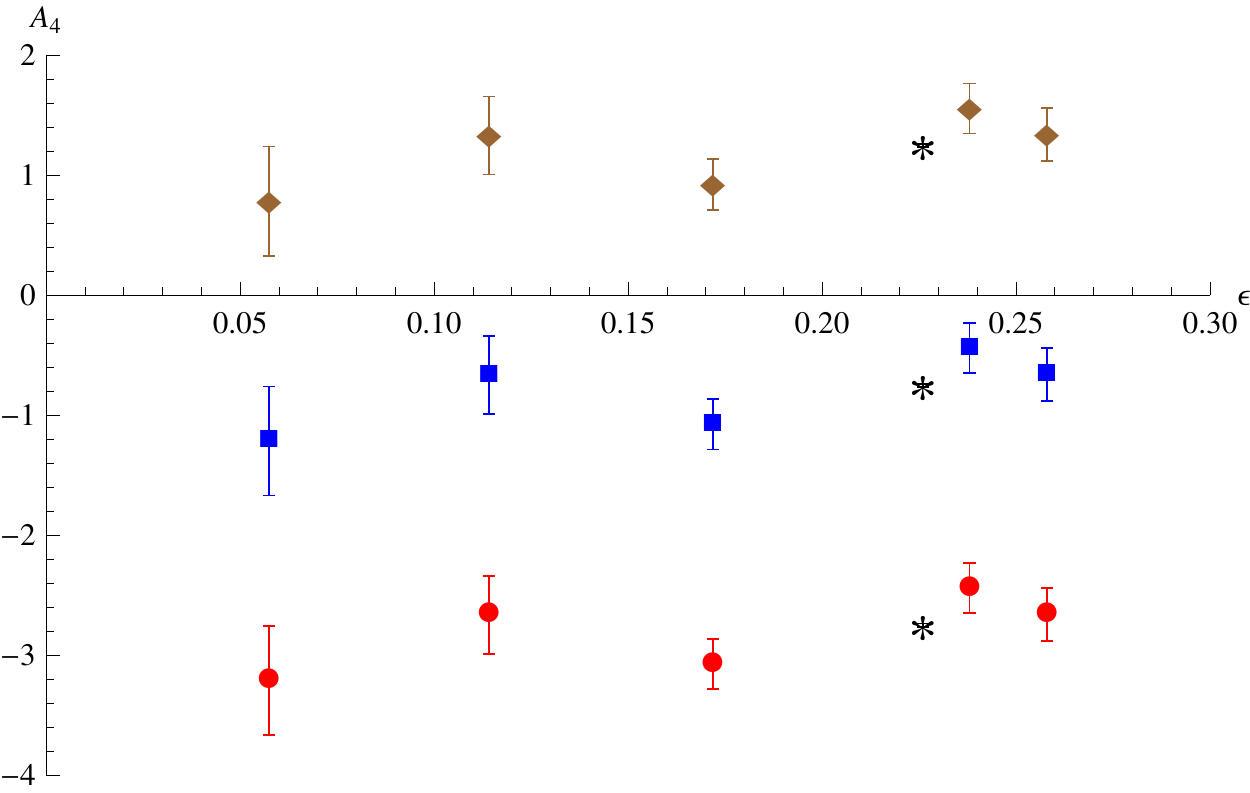}}
  \subfloat[$A_{5}$]{\label{fig:A6Plot}
\includegraphics[width=0.4\textwidth]{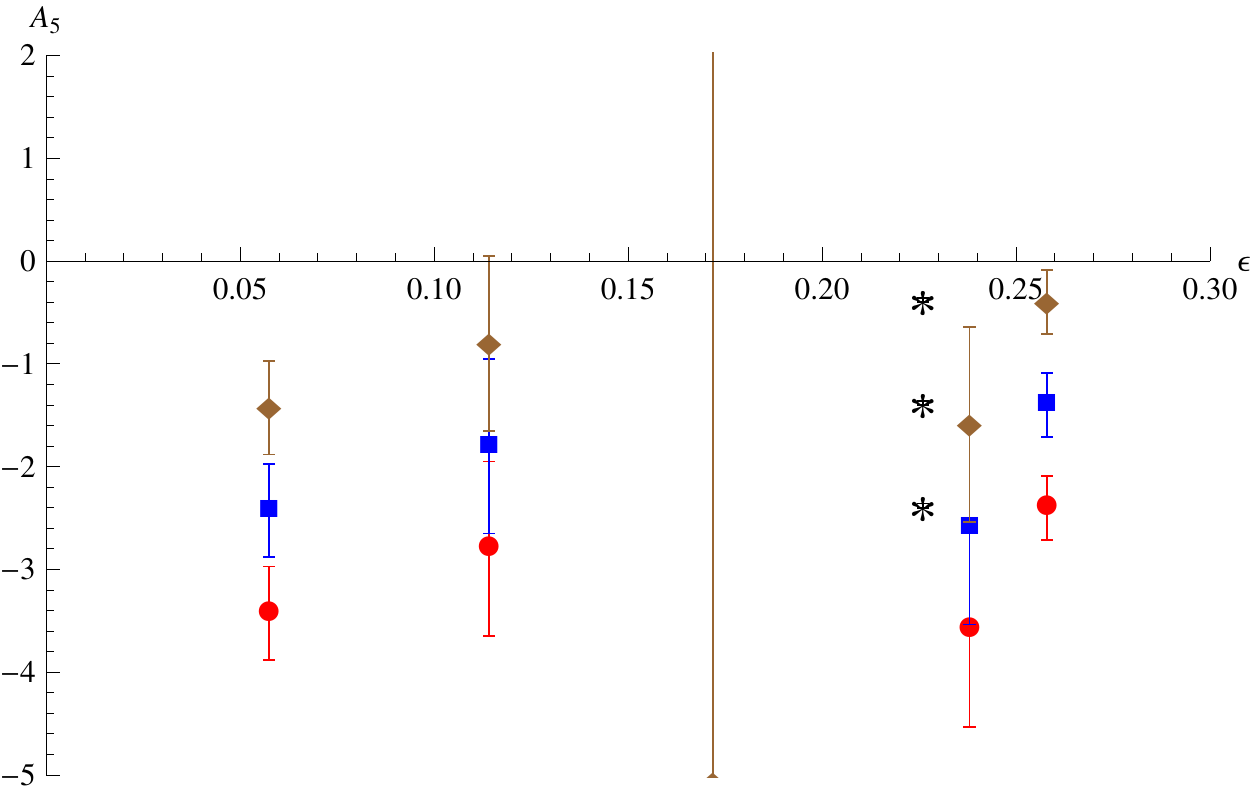}}
  \qquad
  \subfloat[$A_{6}$]{\label{fig:A7Plot}
\includegraphics[width=0.4\textwidth]{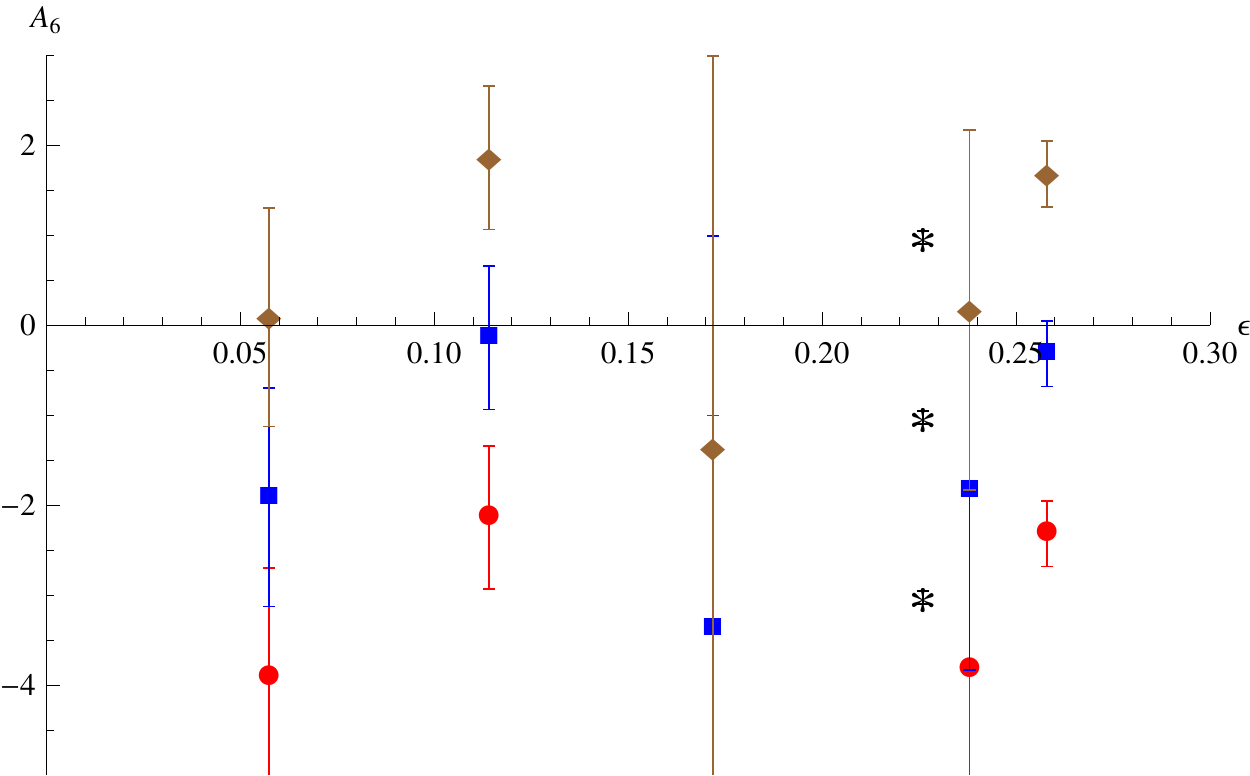}}
  \subfloat[$A_{7}$]{\label{fig:A8Plot}
\includegraphics[width=0.4\textwidth]{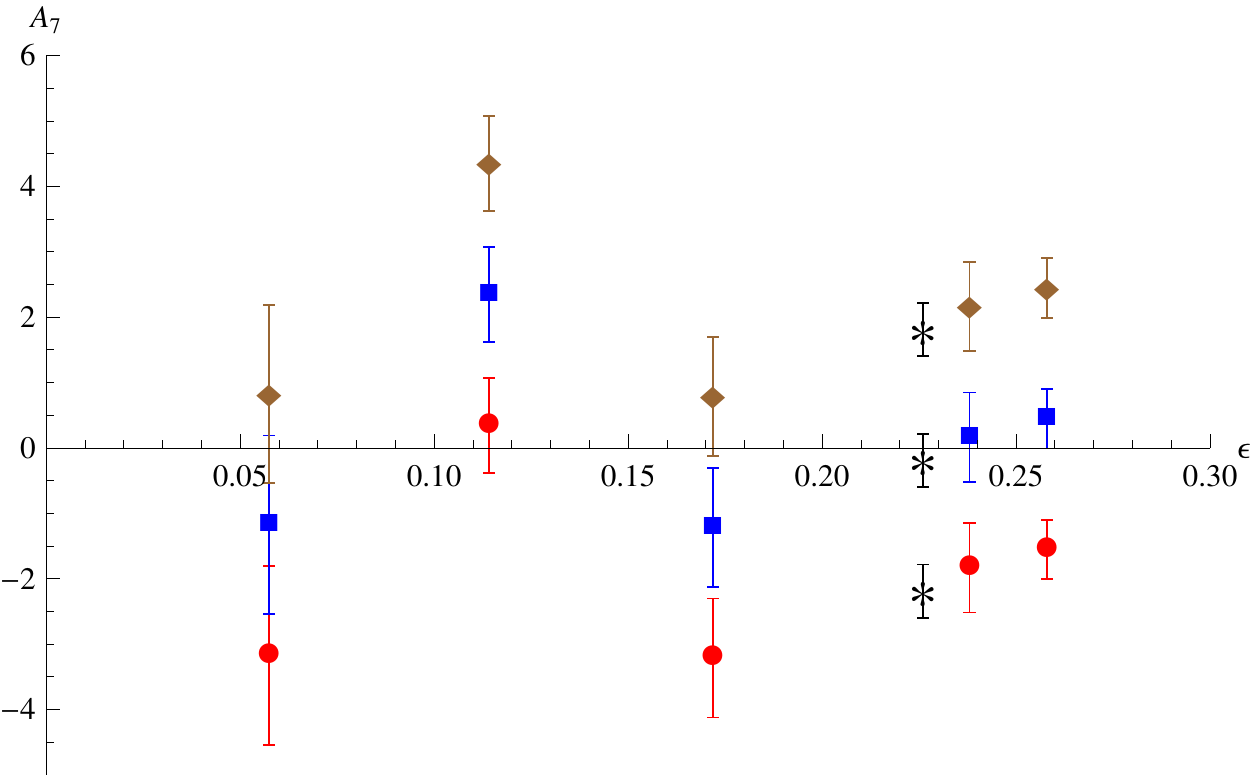}}
  \caption{(Color Online) Plots of the accuracy factors $A_{i}$
  defined in the text.  Black stars denote $A_{i}$ extracted from
  experimental values, red circles are evaluated with an $N_{c}$
  suppression factor of $1$, corresponding to breaking of SU(3)$_{\rm
  flavor}$ only (no $N_{c}$ expansion), blue squares are evaluated
  with $N_{c}^{\rm F}$ suppression factors, and gold diamonds are
  evaluated with $N_{c}^{\rm AS}$ suppression factors. }
  \label{fig:Plots}
\end{figure*}

Figure~\ref{fig:Plots} shows the accuracy factors $A_i$ for each mass
combination as a function of the strength of SU(3)$_{\rm flavor}$
breaking $\epsilon$, along with uncertainties reported in
\cite{Jenkins:2009wv}.  An examination of the plots shows that the
results we found at the physical value of $\epsilon$
in~\cite{Cherman:2009fh} regarding the viability of both large $N_{c}$
limits also apply at generic values of $\epsilon$.  The predictions of
both large $N_{c}$ expansions fit the data equally well ({\it i.e.},
their points cluster in $A_i \in [-1,1]$), and both these large
$N_{c}$ predictions give a much better fit than predictions based
solely on SU(3)$_{\rm flavor}$ breaking.  Furthermore, each $A_i$
appears to possess a smooth $\epsilon \to 0$ limit.  However, evidence
of nonanalytic corrections in $\epsilon$ (which arise in chiral
perturbation theory) has recently been identified in baryon lattice
simulations~\cite{WalkerLoud:2011ab}.

\section{SU(3) symmetry and baryons at large $N_c$}

The existence of lattice calculations for baryon masses with varying
degrees of SU(3) breaking helps to put into stark focus some of the
underlying challenges for the phenomenology of the physical $N_c = 3$
world, since one can now analytically investigate the behavior of
large $N_c$ baryons at more than just a single point.  The crux of the
remaining difficulty is that many more baryon states necessarily
appear at large $N_c$ than at $N_c = 3$.  Accordingly, one needs some
method for relating the baryons at $N_c = 3$ to particular baryons in
the large $N_c$ world.  To keep the discussion focused in this
section, we analyze the situation assuming that the quarks are in the
fundamental representation.  An analogous argument follows for the
case of quarks in the antisymmetric representation, with appropriate
replacements of $N_c^2$ for $N_c$.

For the case of two degenerate flavors, this identification is
obvious.  The standard large $N_c$ analysis~\cite{Gervais:1984rc,
Dashen:1993jt,Dashen:1993as, Dashen:1994qi,Carone:1993dz,Luty:1993fu}
produces ground-state multiplet baryons with $I=J$ differing in mass
at $O(1/N_c)$ [for $J = O(1)$], and with $I=\frac{1}{2}, \frac{3}{2},
\frac{5}{2} \cdots$.  One naturally identifies baryons with
$I=\frac{1}{2}, \frac{3}{2}$ with the analogous $N,
\Delta$ states, respectively, at $N_c = 3$.  Large $N_c$ baryons with
$I \ge \frac{5}{2}$ are taken to be artifacts of the large $N_c$ world
and not relevant at $N_c = 3$.  However, for three degenerate flavors
the analogue of this construction cannot be carried out: {\em All\/}
representations of flavor SU(3) for baryons have dimensions of
$O(N_c^2)$ as $N_c \rightarrow \infty$.  This raises a problem: Which
baryons at large $N_c$ are analogous to ones at $N_c = 3$?  Clearly,
one must declare that almost all of the members in any SU(3) flavor
multiplet are large $N_c$ artifacts, but what principle should one use
to choose?
 
The issue is complicated by the fact that in nature the three flavors
are {\em not\/} degenerate; explicit SU(3) violations occur due to the
mass difference between strange and nonstrange quarks.  One
parametrizes the scale of this SU(3) breaking in the Hamiltonian or in
the meson sector by some dimensionless parameter $\epsilon$.  At
finite but large $N_c$, a typical baryon in an SU(3) multiplet has
$O(N_c)$ strange quarks and a contribution from the SU(3)-violating
term of size $\epsilon N_c$.  As $N_c$ becomes sufficiently large,
$\epsilon N_c \gg 1$, and the notion of ``small violations'' of SU(3)
becomes complicated.  Indeed, one might easily imagine that the
behavior of states might be qualitatively different for varying values
of $\epsilon N_c$.  Thus, one might worry that the behavior of states
at $\epsilon N_c \ll 1$ could be qualitatively different than that for
$\epsilon N_c \sim 1$ or $\epsilon N_c \gg 1$.  The ability to use the
lattice to see how states behave when $\epsilon$ is varied can
therefore play an important role in clarifying the issues.  The fact
that the mass relations of Table~\ref{MassRelations} are quite robust
and work from zero SU(3) breaking to fairly strong breaking provides a
real clue as to what is happening.  As seen below, the explanation is
that the mass relations themselves are quite special in being quite
insensitive to the ambiguities.
 
Let us return to the question of how to identify states at $N_c=3$
with particular states in the large $N_c$ world, and do so without
prejudice from the lattice data.  For simplicity, we consider the case
where $N_c$ is an odd multiple of three.  In principle, one can find
an infinite number of prescriptions to do this.  Two of them are
rather natural: The first one---and the one typically used---is to
identify a baryon at $N_c=3$ with a large $N_c$ baryon of the same
strangeness, total isospin and third component of isospin.  This
identification is particularly natural if $\epsilon N_c$ is
comparatively large, in that one focuses on the lowest-lying states in
the large $N_c$ multiplet.  In effect, one adopts a counting
prescription in $N_s$; the strangeness is treated as being $O(1)$, and
we refer to this identification as the $S$ {\it prescription}.  Such
an approach has a natural analogue in the treatment of topological
solitons such as the Skyrme model, the so-called bound-state approach
of Callan and Klebanov~\cite{Callan:1985hy}.  In that scheme, one
begins by treating a nucleon as an SU(2) Skyrmion and then considers a
strange baryon as a bound state of a kaon and an SU(2) Skyrmion, a
doubly-strange one as two kaons bound to a nucleon, {\it etc.}

However, the approach suggested by the $S$ prescription presents a
difficulty.  Consider the exact SU(3) flavor limit, in which the $u$,
$d$, and $s$ quarks should appear on the same footing, which in turn
means that $I$-spin, $U$-spin and $V$-spin should be treated
equivalently.  This symmetry is badly violated at large $N_c$ by the
identification of states discussed above.  At $N_c=3$, the 18 states
$N$, $\Lambda$, $\Sigma$, $\Xi$, $\Delta$, $\Sigma^*$, $\Xi^*$, and
$\Omega$ associated with the octet and decuplet clearly treat
$I$-spin, $U$-spin, and $V$-spin on the same footing.  For example,
the third component of isospin varies between $-\frac{3}{2}$ and
$+\frac{3}{2}$ for these states, as do the third components of the
$U$-spin and $V$-spin.  Now consider baryons at large but finite
$N_c$, using the $S$ prescription; the third component of isospin
still varies between $-\frac{3}{2}$ and $+\frac{3}{2}$, but the third
components of $U$-spin and $V$-spin are radically different as $N_c$
becomes large: They vary from $(N_c-9)/4$ to $(N_c+3)/4$.  This result
conflicts with the fundamental idea underlying SU(3) flavor symmetry,
that all flavors are created equal.
 
One particular identification of states avoids this problem: Instead
of identifying each state at large $N_c$ with the state at $N_c =3$
of the same strangeness, as in the $S$ prescription, one can identify 
large $N_{c}$ states with $N_{c}=3$ states of the same {\em hypercharge}.  We refer to this as 
the $Y$ {\it prescription}.  Note that, for general $N_c$, the
relationship between hypercharge and number of strange quarks for a
state of baryon number $B$ reads
\begin{equation}
Y=\frac{N_c B}{3} + N_s  \, .
\label{hyper}
\end{equation}
Thus, at large $N_c$ a state with $S = O(1)$ has $Y = O(N_c)$, and
conversely, a state that has $Y = O(1)$ has $S = O(N_c)$.  Using this
new identification, the 18 states of the octet and the decuplet all
have $I_3$, $U_3$, and $V_3$ eigenvalues between $-\frac{3}{2}$ and
$+\frac{3}{2}$.  It is perhaps not surprising that using the
hypercharge to identify states rather than the strangeness does a
better job in preserving the symmetry between $u$, $d$, and $s$ quarks
for this class of states since hypercharge is a traceless generator of
SU(3)$_{\rm flavor}$ but strangeness is not.

The $S$ and $Y$ prescriptions are easy to distinguish in pictorial
form.  In Fig.~\ref{spin32} we exhibit the weight diagram for the
SU(3) representation corresponding to the spin-$\frac 3 2$ baryon
multiplet [the spin-$\frac 1 2$ representation is similar, but has two
sites on the top row and $(N_c+1)/2$ sites on the long sides].  In the
$S$ prescription, the analogues to the $N_c = 3$ baryons appear in the
top rows (minimum $N_s$) of the diagram, while in the $Y$ prescription
they coincide with the sites nearest the centroid ($Y = 0$) of the
diagram.

\def\onedot{\makebox(0,0){$\scriptstyle 1$}}
\def\twodot{\makebox(0,0){$\scriptstyle 2$}}
\def\threedot{\makebox(0,0){$\scriptstyle 3$}}
\def\fourdot{\makebox(0,0){$\scriptstyle 4$}}

\setlength{\unitlength}{3mm}

\begin{figure}
  \begin{centering}
\centerline{\hbox{
\begin{picture}(20.79,18)(-8.085,-8)
\multiput(-1.155,10)(2.31,0){4}{\onedot}
\multiput(-2.31,8)(9.24,0){2}{\onedot}
\multiput(-3.465,6)(11.55,0){2}{\onedot}
\multiput(-4.62,4)(13.86,0){2}{\onedot}
\multiput(-5.775,2)(16.17,0){2}{\onedot}
\multiput(-6.93,0)(18.48,0){2}{\onedot}
\multiput(-8.085,-2)(20.79,0){2}{\onedot}
\multiput(-6.93,-4)(18.48,0){2}{\onedot}
\multiput(-5.775,-6)(16.17,0){2}{\onedot}
\multiput(-4.62,-8)(2.31,0){7}{\onedot}
\multiput(0,8)(2.31,0){3}{\twodot}
\multiput(-1.155,6)(6.93,0){2}{\twodot}
\multiput(-2.31,4)(9.24,0){2}{\twodot}
\multiput(-3.465,2)(11.55,0){2}{\twodot}
\multiput(-4.62,0)(13.86,0){2}{\twodot}
\multiput(-5.775,-2)(16.17,0){2}{\twodot}
\multiput(-4.62,-4)(13.86,0){2}{\twodot}
\multiput(-3.465,-6)(2.31,0){6}{\twodot}
\multiput(1.155,6)(2.31,0){2}{\threedot}
\multiput(0,4)(4.62,0){2}{\threedot}
\multiput(-1.155,2)(6.93,0){2}{\threedot}
\multiput(-2.31,0)(9.24,0){2}{\threedot}
\multiput(-3.465,-2)(11.55,0){2}{\threedot}
\multiput(-2.31,-4)(2.31,0){5}{\threedot}
\multiput(2.31,4)(2.31,0){1}{\fourdot}
\multiput(1.155,2)(2.31,0){2}{\fourdot}
\multiput(0,0)(2.31,0){3}{\fourdot}
\multiput(-1.155,-2)(2.31,0){4}{\fourdot}
\end{picture}
}}

\bigskip

\caption{Weight diagram for the ground-state spin-$\frac 3 2$
representation of SU(3), adapted from Ref.~\cite{Dashen:1994qi}.  The
numbers indicate the multiplicity of states at a given site.  While
this literal diagram corresponds to $N_c = 15$, it is easily
generalized to arbitrary odd $N_c$ by extending the longer sides to
$(N_c-1)/2$ sites.}
\label{spin32}

  \end{centering}
\end{figure}
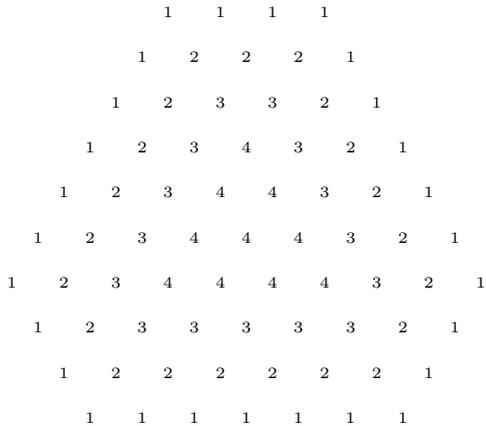

The two prescriptions have important physical differences.  If one
focuses on the mass of a typical baryon $M$ in the octet or decuplet
and considers its dependence on the three quark masses, one finds
\begin{align}
&{\rm   Prescription} \,  S: \nonumber\\
&\frac{d M}{d m_s} = N_s = O(N_c^0) \, , \; \; \; \; \;\; \;
\frac{d M}{d m_{u,d}} = N_{u,d} = O(N_c^1) \, , \nonumber \\
& {\rm   Prescription} \, Y: \nonumber\\
&\frac{d M}{d m_{u,d,s}} = N_{u,d,s} = O(N_c^1) \, .
 \; \;\end{align}
Note that the rate of change of the baryon mass with $m_s$ is
qualitatively different in the two approaches.  Since this rate also
equals $\langle B|\int {\rm d}^3 x \, \bar{s} s | B
\rangle$, the scalar strangeness content of the baryon behaves
differently in the two approaches, while $\langle B|\int {\rm
d}^3 x \, \bar{u} u | B \rangle \sim \langle B|\int {\rm d}^3 x \,
\bar{d} d | B \rangle = O(N_c)$ in either approach.  Given that the
two prescriptions are physically quite different at large $N_c$, it
should be clear that they actually correspond to distinct $1/N_c$
expansions for the baryon masses.  One then faces the obvious
question: Which one of these expansions is more phenomenologically
useful in describing the world of $N_c = 3$?

It is noteworthy that Ref.~\cite{Jenkins:1995td}, where the mass
relations of Table~\ref{ops_S_Y} were first derived, makes the
explicit assumption that the strangeness --- rather than the
hypercharge --- is of order unity, which is consistent with prescription
$S$.  The fact that the lattice suggests the $N_c = 3$ mass relations
are robust---holding qualitatively over a fairly wide ride range of
the scale of SU(3) breaking $\epsilon$, may then seem to suggest that
prescription $S$ is the phenomenologically relevant choice.  Such a
result may seem a bit puzzling, since it is plausible that
prescription $Y$, which unlike prescription $S$ treats $u$, $d$, and
$s$ on an equal footing, is more sensible than prescription $S$ very
near the SU(3) limit.

However, this result produces no real puzzle.  The robustness of the
mass relations tells us more about the nature of the relations
themselves than about which prescription is better.  Except for
$M_0$, the relations all involve the {\em differences\/} of masses
rather than masses themselves, and while the individual masses are
sensitive to the choice of prescription, we argue that this
sensitivity cancels completely in the mass relations to the order that
the relations hold.  Thus, the relations are themselves quite robust,
and this fact appears to be responsible for the relations holding at
the predicted level of accuracy, even for widely varying values of
SU(3) breaking.  Additional evidence for the robustness of the mass
analysis under an alternate treatment of the flavor quantum numbers
appears in Ref.~\cite{Bedaque:1995wb}.

To see how the cancellations arise, recall from Sec.~\ref{sec:Baryons}
how the mass relations are derived from operator relations.  Assuming
exact isospin symmetry and using quarks in the fundamental
representation gives the mass Hamiltonian in Eq.~(\ref{ops}), whose
operators have matrix elements given in Eq.~(\ref{matrixelements}).
Using these expressions, the mass Hamiltonian can be recast in the
equivalent two forms:
\begin{equation}
M =  \sum_i a_i^S \, O_i^S =  \sum_i a_i^{Y} \,  O_i^{Y} \, ,
\label{opsform}
\end{equation}
where the $a$ coefficients are linear combinations of the $c$
coefficients and can be shown to be of order unity; the operators are
defined in Appendix~\ref{newbases} to have the simple matrix elements
given in Table~\ref{ops_S_Y}.  For simplicity, the leading $N_c$
dependence and $\epsilon$ dependence are included as part of the
operators $O^S_i$ and $O^Y_i$.  The superscripts $S$ and $Y$ indicate
which of the two prescriptions is used in the $N_c$ counting.

The two sets of operators have one very important property: All
operators of type $S$ at a given order in $\epsilon$ and
$\frac{1}{\N}$ can be written as linear combinations of operators of
type $Y$ at equal or subleading order in both $\epsilon$ and
$\frac{1}{\N}$.  Thus, for example, $O_4^S$, whose matrix elements are
$O(\epsilon/N_c)$, is expressible in terms of $O_1^{Y}$, $O_3^{Y}$,
and $O_4^{Y}$, which are all of equal or subleading order in both
$\epsilon$ and $\frac{1}{\N}$.  It is straightforward to explicitly
show that this property holds for all of the operators, and this
result essentially amounts to a statement about the decomposition of
any high-order tensor into an irreducible piece plus subleading
traces.  Similarly, all operators of type $Y$ at a given order in
$\epsilon$ and $\frac{1}{\N}$ can be written as linear combinations of
operators of type $S$ at equal or subleading order in both $\epsilon$
and $\frac{1}{\N}$.

Let us now recall how the mass relations are obtained.  One finds
linear combinations of the masses among the eight independent baryons $N$,
$\Lambda$, $\Sigma$, $\Xi$, $\Delta$, $\Sigma^*$, $\Xi^*$, and
$\Omega$ that vanish when acted on by all operators of an equal or
subleading order in both $\epsilon$ and $\frac{1}{\N}$.  Each such
combination gives a mass relation that holds up to the given order in
$\epsilon$ and $\frac{1}{\N}$.  However, since any operator of type
$S$ can be written as a superposition of operators of type $Y$ at
equal and subleading order, any mass relation that holds for
prescription $S$ also holds for prescription $Y$, and vice versa.

\begin{widetext}
\begin{center}
\begin{table}
\caption{The eight operators given in Eq.~(\ref{opsform}), including the
$N_c$ and $\epsilon$ dependence.  Prescriptions $S$ and $Y$ are
defined in the text.
\label{ops_S_Y}}
\begin{tabular}{lccccccl}
&&Operator && Prescription $S$ && Prescription $Y$\\ \hline\hline
&&$O_1$&& $\N$ && $\N$ \\
\hline
&&$O_2$ && $\frac{1}{\N} J(J+1) $&&$\frac{1}{\N} J(J+1) $\\
\hline
&&$O_3$ && $\epsilon \,  N_s $&&$-\epsilon \, Y$\\
\hline
&&$O_4$ && $ \frac{\epsilon}{\N} \, \left[ I(I+1)- \frac{N_s}{2}
\left( \frac{N_s}{2} +1 \right) \right] $&& $\frac{\epsilon}{\N} \,
\left[ I(I+1) - \frac{Y}{2}\left (\frac{Y}{2} -1 \right) \right] $ \\
\hline
&&$O_5$ && $\frac{\epsilon}{\N^2} \, N_s  \, J(J+1)$&&$-
\frac{\epsilon}{\N^2}\, Y  \, J(J+1)$\\
\hline
&&$O_6$ && $\frac{\epsilon^2}{\N} \, N_s^2$ && $\frac{\epsilon^2}{\N}
\,Y^2$\\
\hline
&&$O_7$ && $\frac{\epsilon^2}{\N^2} \, N_s \,  \left[ I(I+1)-
\frac{N_s}{2}\left (\frac{N_s}{2} +1 \right) \right] $&&
$-\frac{\epsilon^2}{\N^2} \, Y \, \left[ I(I+1)- \frac{Y}{2}
\left( \frac{Y}{2} -1 \right) \right]$\\
\hline
&&$O_8$ && $\frac{\epsilon^3}{\N^2} \, N_s^3$ && $-
\frac{\epsilon^3}{\N^2} \,Y^3$\\
\hline
\hline
\end{tabular}
\end{table}
\end{center}
\end{widetext}

In terms of Eq.~(\ref{opsform}), our result indicates that $a_i^S =
a_i^Y +$ equal or subleading order in $\epsilon$ and $1/N_c$.  One
additional point deserves mention: If the correction is of equal
order, one might fear that only one of the two prescriptions gives
fully hierarchical coefficients, the other one merely ``maintaining
the status quo.''  In the example of $O_4$ given above, $O_4^S = O_4^Y
- \frac 1 6 O_3^Y$ + subleading order, so one might expect that either
$a_4^S$ is no smaller than $a_3^S$ or $a_4^Y$ is no smaller than
$a_3^Y$, at least parametrically.  But in fact, one can check that the
numerical hierarchy using the physical baryon masses supports the
result that both $a_4$'s are $O(N_c) \sim 3$ smaller than their
corresponding $a_3$'s, owing to the smallness of the explicit $-\frac
1 6$ coefficient.

To summarize, individual masses depend sensitively on one's choice of
prescription, but the mass relations do not.  They hold at the stated
order of accuracy regardless of how one chooses to identify baryons at
large $N_c$ with $N_c=3$ baryons.  In particular, the validity of the
mass relations does not depend upon the scale of $\epsilon N_c$;
rather they only depend on both $\epsilon$ and $1/N_c$ to be
separately small.  The fact that lattice results uphold the mass
relations at the expected levels of accuracy, even as $\epsilon N_c$
varies widely, presumably reflects this fact.

Ultimately, this behavior might teach us an important lesson about the
applicability of large $N_c$ operator analysis to baryons with three
flavors.  To find quantities that discriminate between
possible prescription choices, one presumably needs to consider
quantities sensitive to the absolute number of quarks of various types
in a given state, rather than their relative number between different
baryon states.  The one place in our analysis that this distinction
might be possible is the common mass accuracy parameter $A_0$.
However, a linear fit to the points in Fig.~\ref{fig:Plots}(a) gives
$A_0 = 0.215 - 0.177 \epsilon$.  Since 0.177/0.215 is neither large
nor small compared to unity, one cannot distinguish decisively between
$M_0 \sim N_c + \epsilon$ ($S$ prescription) or $M_0 \sim N_c +
\epsilon N_c$ ($Y$ prescription).

\label{sec:NcandSU3}

\section{Conclusions}
\label{sec:Conclusions}

Lattice simulations provide a unique window into understanding nonperturbative physics and why our universe chooses one unique solution out of many possibilities, particularly since the simulations allow one to explore universes in which the underlying parameters (quark masses, for example) can be chosen at will.   Another very recent example studies
baryon masses by varying the literal numerical value of
$N_c$~\cite{DeGrand:2012fk}.  Such results provide otherwise unattainable
insights into strongly interacting systems.

In this work we have seen that lattice simulations of baryons over a
range of SU(3)$_{\rm flavor}$-breaking parameter $\epsilon$ provide a
spectrum of masses not only explainable in terms of a large $N_c$ QCD
expansion (as seen in previous work), but are in fact agnostic as to
whether the quarks fill the fundamental or two-index antisymmetric
representation of SU($N_c$).  Moreover, either expansion does a much
better job accounting for the mass spectrum than including only the
$\epsilon$ dependence and ignoring all factors of $N_c$.  This result
greatly extends the scope of our previous work, which reached the same
conclusion but only at the physically realized value of $\epsilon$.

We also addressed the interesting question of why this conclusion
should hold as $\epsilon \to 0$, the SU(3)-symmetric point, when the
baryon states in these analyses are assumed to have $O(1)$ strange
quarks and $O(N_c)$ $u$ and $d$ quarks---a highly SU(3)-asymmetric
configuration.  The resolution appears to be that mass differences,
which constitute the bulk of the large $N_c$ baryon results, are
insensitive to whether one works in a prescription that favors
minimizing either the number of $s$ quarks ($S$~prescription) or the
difference between the number of $u$, $d$, and $s$ quarks
($Y$~prescription).  An examination of observables sensitive not to
differences of quarks but rather to their collective effect is necessary
to resolve this remarkable ambiguity.

\begin{acknowledgments}
The authors thank the organizers of CAQCD-2011 (U.~Minnesota), at
which the initial discussions inspiring this work were held.  T.D.C.\
acknowledges the support of the U.S.\ Department\ of Energy under Grant
No.\ DE-FG02-93ER-40762.  R.F.L.\ acknowledges the support of the NSF
under Grants No.\ PHY-0757394 and No.\ PHY-1068286, and thanks the Galileo
Galilei Institute for Theoretical Physics for their hospitality and
the INFN for partial support.
\end{acknowledgments}

\appendix
\section{Relationships between operator bases}
\label{newbases}

The precise relationships between the operators given in
Eq.~(\ref{matrixelements}) and the operators $O^{S,Y}_i$ of the $S$
and $Y$ prescriptions that give the matrix elements listed in
Table~\ref{ops_S_Y} are:
\begin{eqnarray}
O^{S,Y}_1 & = & O^{1,0}_{(0)} \, , \nonumber \\
O^{S,Y}_2 & = & O^{1,0}_{(2)} \, , \nonumber \\
O^S_3 & = & -\frac{2}{\sqrt{3}} O^{8,0}_{(1)} + \frac{\epsilon}{3}
O^{1,0}_{(0)} \, , \nonumber \\
O^Y_3 & = & -\frac{2}{\sqrt{3}} O^{8,0}_{(1)} \, , \nonumber \\
O^S_4 & = & \frac{2}{\sqrt{3}} O^{8,0}_{(2)} + \frac{\epsilon}{3}
O^{1,0}_{(2)} \, , \nonumber \\
O^Y_4 & = & \frac{2}{\sqrt{3}} O^{8,0}_{(2)} + \frac{\epsilon}{3}
O^{1,0}_{(2)} + \frac{\epsilon}{6} \left( \frac 1 6 + \frac{1}{N_c}
\right) O^{1,0}_{(0)} \nonumber \\
& & -\frac{1}{3\sqrt{3}} O^{8,0}_{(1)} \, ,
\nonumber \\
O^S_5 & = & -\frac{1}{\sqrt{3}} O^{8,0}_{(3)} + \frac{\epsilon}{3}
O^{1,0}_{(2)} \, , \nonumber \\
O^Y_5 & = & -\frac{1}{\sqrt{3}} O^{8,0}_{(3)} \, , \nonumber \\
O^S_6 & = & \frac 2 3 O^{27,0}_{(2)} -\frac{4\epsilon}{3\sqrt{3}}
O^{8,0}_{(1)} +\frac{\epsilon^2}{9} O^{1,0}_{(0)} \, ,
\nonumber \\
O^Y_6 & = & \frac 2 3 O^{27,0}_{(2)} \, , \nonumber \\
O^S_7 & = & -\frac 2 3 O^{27,0}_{(3)} -\frac{\epsilon}{3\sqrt{3}}
O^{8,0}_{(3)} +\frac{2\epsilon}{3\sqrt{3}} O^{8,0}_{(2)} +
\frac{\epsilon^2}{9} O^{1,0}_{(2)} \, , \nonumber \\
O^Y_7 & = & -\frac 2 3 O^{27,0}_{(3)} + \frac 1 9 O^{27,0}_{(2)}
-\frac{\epsilon}{3\sqrt{3}} O^{8,0}_{(3)} \nonumber \\
& & - \left( \frac 1 6 + \frac{1}{N_c}
\right) \frac{\epsilon}{3\sqrt{3}} O^{8,0}_{(1)}\, , \nonumber \\
O^S_8 & = & -\frac{2}{3\sqrt{3}} O^{64,0}_{(3)} + \frac{2 \epsilon}{3}
O^{27,0}_{(2)} -\frac{2\epsilon^2}{3\sqrt{3}} O^{8,0}_{(1)} +
\frac{\epsilon^3}{27} O^{1,0}_{(0)} \, , \nonumber \\
O^Y_8 & = & \, -\frac{2}{3\sqrt{3}} O^{64,0}_{(3)} \, .
\end{eqnarray}

\bibliography{LargeNBaryons}
  
\end{document}